# Complexity Metrics for Spreadsheet Models

*Andrej Bregar*
*University of Maribor*
*Faculty of Electrical Engineering and Computer Science*
*Smetanova 17, SI-2000 Maribor, Slovenia*
*Email: andrej.bregar@uni-mb.si*

**ABSTRACT**

*Several complexity metrics are described which are related to logic structure, data structure and size of spreadsheet models. They primarily concentrate on the dispersion of cell references and cell paths. Most metrics are newly defined, while some are adapted from traditional software engineering. Their purpose is the identification of cells which are liable to errors. In addition, they can be used to estimate the values of dependent process metrics, such as the development duration and effort, and especially to adjust the cell error rate in accordance with the contents of each individual cell, in order to accurately asses the reliability of a model. Finally, two conceptual constructs – the reference branching condition cell and the condition block – are discussed, aiming at improving the reliability, modifiability, auditability and comprehensibility of logical tests.*

## 1 INTRODUCTION

Many spreadsheet models are large and complex. Because they are rarely built according to formal software analysis and design strategies, and because cell formulas are hidden, most of these models contain serious errors. Several research studies of fault rates have been performed in the past [Panko, 2000; Panko and Sprague, 1998], indicating that 20 to 60 percent of spreadsheets produce wrong outputs. For this reason, minimization of errors through proper design, quality assurance and systematic cell inspection is crucial. Some structured development/modelling methods have already been applied in order to reduce risks [Conway and Ragsdale, 1997; Janvrin and Morrison, 2000; Kreie et al., 2000; Read and Batson, 1999; Ronen et al., 1989]. Moreover, useful auditing techniques and tools exist, which assist developers in testing and maintaining spreadsheet models, as well as in understanding their structure [Clermont and Mittermeir, 2003; Mittermeir and Clermont, 2002; Nixon and O'Hara, 2001].

In this paper, several metrics are defined, which aim at estimating the complexity of spreadsheet models. It has been proven that the complexity of a model or a particular formula represents an important factor to be considered in the process of spreadsheet development, because a complex spreadsheet makes error finding difficult [Teo and Lee-Partridge, 2001] and because errors come in relation with cells that have a high potential for faults [Panko, 2000]. Various product and process metrics have been extensively used in the software engineering field for a couple of decades [Conte et al., 1986]. Some of them may be adopted for the purpose of spreadsheet development, yet new ones should be introduced in order to detect cells that are especially prone to errors.

It should be stressed that this paper presents work in progress. Hence, the metrics defined here are not exhaustive, and they have neither been validated in practice. The paper is organized as follows. In Section 2, the complexity metrics are discussed. Section 3 gives an example to show why it is sensible to adjust the cell error rate in accordance with the estimated degree of complexity, in order to accurately asses the level of bottom-line reliability. Since the definition of logic structure metrics indicates that spreadsheets do not permit efficient modelling of complex logical tests, two conceptual constructs are proposed in Section 4 – the reference branching condition cell and the condition block. Directions for further work are stated in the concluding Section 5.

## 2 COMPLEXITY METRICS

Figure 1 gives a summary of metrics. The formula size and structure metrics are mostly adapted from traditional software engineering. They comprise measurements of logic structure, because tests and branches are possible only within individual formulas. Since there can be no loops, a sensible metric is the *decision count*. It is the number of simple conditions – conjunction and disjunction predicates – within one formula.

| *Formula size* | *Cell range* |
|---|---|
| - Number of operators<br>- Number of operands | - Range width (height) |
| *Formula structure* | *Cell cascade* |
| - Nesting level of a token<br>- Average nesting level<br>- Depth of nesting<br>- Decision count | - Cell fan-in<br>- Cell fan-out<br>- Reachability of a cell<br>- Average reachability<br>- Average path length<br>- Maximal path length<br>- Total number of paths |
| *Cell references of a formula* | *Modular structure* |
| - Dispersion of references<br>- Single column (row) reference delta<br>- Maximal positive (negative) delta | - Number of data binding triples<br>- Percentage of unreferenced data |

Figure 1: Summary of metrics with regard to usage

An important complexity factor is represented by the nesting levels of formula operators and operands. Nesting occurs, because each function operand can be a result of another function. Two metrics can be defined: the *depth of nesting* and the *average nesting level*. The latter is computed in the following way:

$$NL_{Avg} = \frac{\sum_{i=1..N_1+N_2} NL_i}{N_1 + N_2}.$$

Here, $N_1$ and $N_2$ are the numbers of total occurences of formula operators and operands, and $NL_i$ is the nesting level of *i*-th token. Literal values and variables (cell references) are considered as operands. Operators include function names and special symbols. It should be stressed that the depth of nesting in a formula can be reduced by assigning primitive functions to other cells in a worksheet and making references to those cells. In this case, however, the lengths of cell cascades increase. Consequently, unnecessary computational details are exposed, and it becomes more difficult to put input data in close physical proximity to dependent results of calculations. It therefore remains an interesting task to prove experimentally whether reducing nesting improves auditability and modifiability of a spreadsheet. Another point to notice is that the number of all cell references increases. Yet, these references are rather uniformly distributed among several formulas, implying that the average number of references per single cell lessens. Thus, a correlated research question worth investigating is whether there exists a higher probability of linking errors when there are $N$ references altogether and averagely $n$ references in a single cell, or when there are $M$ respectively $m$ references, so that $N < M$ and $n > m$. According to some studies [Janvrin and Morrison, 2000; Panko, 2000; Teo and Lee-Partridge, 2001], linking errors are classified as a special case of mechanical spreadsheet errors. They represent wrong cell references which are destroying the integrity of subsequent modules, variables and algorithms.

Because there can be many ways to reach each formula cell in a cascade, a useful metric is the *reachability* of a cell. To define it, the number of direct links that lead in a cell and the number of direct links out of a cell have to be measured. *Fan-in* is thus the count of references to other cells (*precedents*), while *fan-out* is the count of references from other cells (*dependents*). It can now be said that a bottom-line cell is a cell with fan-out equal to zero. It has no dependents, and it is considered to be an output variable. It terminates a computational cascade which can have several independent zero-fan-in starting cells with no precedents. These cells store data, and are regarded as input variables.

One approach to determine reachability is shown on Figure 2. It describes each path as a sequence of arcs from one of the start cells to the relevant terminal cell.

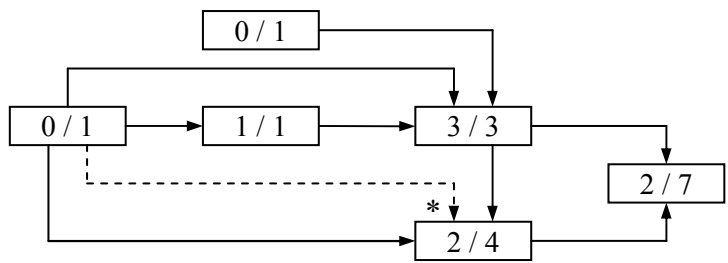

Figure 2: Paths through a cell cascade

The first number written within each symbol on Figure 2 represents the fan-in of a cell, while the second number denotes its reachability. It can be seen from the example that there are four unique paths to reach the cell marked with an asterisk. Its reachability is hence equal to four. Similarly, the reachability of the bottom-line result cell is seven. It is fairly evident that the reachability of a cell is computed by summing the reachabilities of preceding cells that are directly linked by references, unless a cell has no precedents, in which case its reachability equals to one:

$$R(C) = \begin{cases} 1 & , FI(C) = 0 \\ \sum_{i=1..FI(C)} R(C_i) & , \text{otherwise.} \end{cases}$$

In this formula, $R(C)$ is the reachability and $FI(C)$ fan-in of the observed cell $C$. Cells, which are referenced by $C$, are denoted with $C_i$, $i = 1, \ldots, FI(C)$; their reachabilities are marked with $R(C_i)$.

Conceptually, each computational cascade has only one terminal zero-fan-out cell. The total number of paths in a cascade equals the reachability of this bottom-line cell. The average reachability of a cascade is easily calculated; $R_{Avg}$ for the given example is 17/6. Also, the number of cells in each path can be counted, the path with the maximum length can be found, and the average length can be computed.

Cell ranges should be given special treatment. Firstly, a reference to a range has to be transformed into as many single-cell references as there are cells in a range. This is required because each cell is accessed and processed separately. Thus, each cell in a range is on its own independent path. Secondly, a series of calculations in a range with copied formulas could be regarded as a loop with a fixed number of iterations. It tends to be more auditable than a group of cells with different formulas, but it exhibits a risk potential because of possible invalid absolute and relative references. Two size metrics – the *width* ($S_W$) and the *height* ($S_H$) of a range – are therefore defined. Suppose that there are two single-column ranges, $A$ and $B$. Cell formulas in the range $B$ reference cells in the range *A*, so that *s* successive cells are accessed at a time. The equation $S_{HA} = s$ must hold true in the case of absolute referencing, regardless of the size $S_{HB}$, and $S_{HA} = S_{HB} + s - 1$ must hold true when linking *A* and *B* via relative references. Otherwise, there exists a great probability of error.

It has been shown that mechanical error rates rise dramatically when equations contain references to cells that are in both different columns and different rows than a cell containing the formula [Panko, 2000]. For this reason, the *dispersion of references* is measured. The most simple way to define it is to use the exponential utility function:

$$DR = 1 - \exp(-\alpha \cdot \Delta),$$
$$\Delta = \sum_{i=1..N} | DX_i \cdot DY_i |.$$

Here, *N* is the number of formula operands representing cell references, $DX_i$ and $DY_i$ are the *column* and *row reference deltas* for the *i*-th cell reference and *α* is a small positive constant. If a referenced cell is in the same column or row as the formula cell, then either $DX_i$ or $DY_i$ equals to zero, and the reference does not contribute to the dispersion. The exponential function is applied for the computation of *DR* because cell distances do not linearly influence the model readability; however, the sigmoid function form could be used as well. The constant *α* sets the slope of the dispersion function and determines which distance sums are "still good enough" in terms of nearly all quality characteristics, including the communicative effectiveness, comprehensibility and design. The value of *α* has an order of magnitude of $10^{-2}$ and has been set experimentally. Figure 3 demonstrates that pretty realistic degrees of dispersion are obtained for $\alpha = 0.01$.

| Δ | DR |
|---|---|
| 10 | 0.0952 |
| 20 | 0.1813 |
| 50 | 0.3935 |
| 100 | 0.6321 |
| 150 | 0.7769 |
| 200 | 0.8647 |
| 300 | 0.9502 |

Figure 3: Dispersions of references obtained for $\alpha = 0.01$

Other equations to determine the dispersion degree will have to be defined and evaluated within the scope of further research work, as the proposed formula has several drawbacks.

1. Even if the referenced cell is in the same column or row (either $DX_i = 0$ or $DY_i = 0$) as the observed cell formula, the non-zero reference delta has to influence the calculated level of dispersion. Especially, if a co-ordinate delta is large (for example, when the Manhattan distance exceeds 20 cells), such a reference hinders the ability to audit, test, modify and comprehend a spreadsheet model. To take into account this kind of reference deltas, the $|DX_i \cdot DY_i|$ product will be replaced by $L_1$ (Manhattan) and $L_2$ (Euclidean) distances.
2. If, simultaneously, cells in the same column and cells in the same row are referenced, so that $\exists\ i, j: DX_i = 0 \land DY_i \neq 0 \land DX_j \neq 0 \land DY_j = 0$, the auditability, modifiability, comprehensibility and other quality characteristics additionally deteriorate.
3. References may be balanced or unbalanced. In the first case, they are all oriented to the common direction, which means that they point to cells which are either close to a single co-ordinate axis or which are in the same quadrant. For this reason, it would be sensible to consider references as two-dimensional vectors, and to determine angles between these vectors. Clearly, this would be quite a restrictive approach, since most modellers obey good spreadsheet design practice which advises us to make references solely to the left or upwards [Read and Batson, 1999].

In order to partially overcome the three stated problems, the dispersion of references is supplemented with the *column span* and *row span*. Spans are computed by subtracting maximal negative reference deltas from maximal positive deltas. These measures are necessary, because a matrix of approximately twenty times twenty cells is visible to the spreadsheet modeller, auditor, user or maintainer. Large spans therefore noticeably reduce the auditability and reliability. The dispersion does not carry this information, as it varies according to the number of references, and may distort the complexity picture, if some referenced cells are in the same column and some other in the same row as the treated formula.

A way to improve spreadsheet design is the structuring approach [Janvrin and Morrison, 2000]. Large models are modularized by creating worksheets. Each individual worksheet becomes a structured module that takes a defined set of inputs and translates them into a prescribed set of outputs. Values are linked as outflows and inflows from one module to another by cell references. Each major component of a spreadsheet is thereby represented as a different module, and all these modules are integrated into a coherent whole.

If a spreadsheet has such a modular design, the sharing of data among modules can be measured. One approach, which is adopted from traditional software engineering [Conte et al. 1986], is to count the number of *data binding triples (P, Q, R)*. The variable $Q$ is set by the module $P$ and read by the module $R$. The more triples there are for the modules $P$ and $R$, the more complex is their relationship. All variables in a spreadsheet model are global; they can be accessed from anywhere, but they do not necessarily pass from one module to another. The lower fan-in and fan-out the module has, the more independent it is. This certainly has an important influence on reliability, modifiability and auditability.

## 3. ESTIMATION OF BOTTOM-LINE ERROR RATE

In programming, one of the basic metrics is the number of faults per thousand lines of non-comment code. A similar metric for spreadsheet models is the *cell error rate* (*CER*) – the percentage of non-label cells containing errors [Panko and Sprague, 1998]. CER is estimated to be between 1% and 2%. However, according to conducted research studies, between 20% and 60% of spreadsheet models contain at least one serious error [Panko, 2000]. The difference is due to the fact that bottom-line values are computed through long cascades of formula cells. Because in tasks that contain many sequential or subordinate operations error rates multiply along cascades of subtasks, the fundamental equation for computing the *bottom-line error rate* is based on a memoryless geometric distribution over cell errors:

$$E = 1 - (1 - e)^n.$$

Here, $E$ is the bottom-line error rate, $e$ is the cell error rate and $n$ is the number of cells in the cascade. $E$ indicates the probability of an incorrect result in the last cascade cell, given the probability of an error in each cascade cell is equal to the cell error rate.

Figures 4 and 5, respectively, provide an example to show that the cell error rate should necessarily be adjusted in accordance with the complexity of each individual cell. This is a prerequisite for the bottom-line error rate to be accurately estimated.

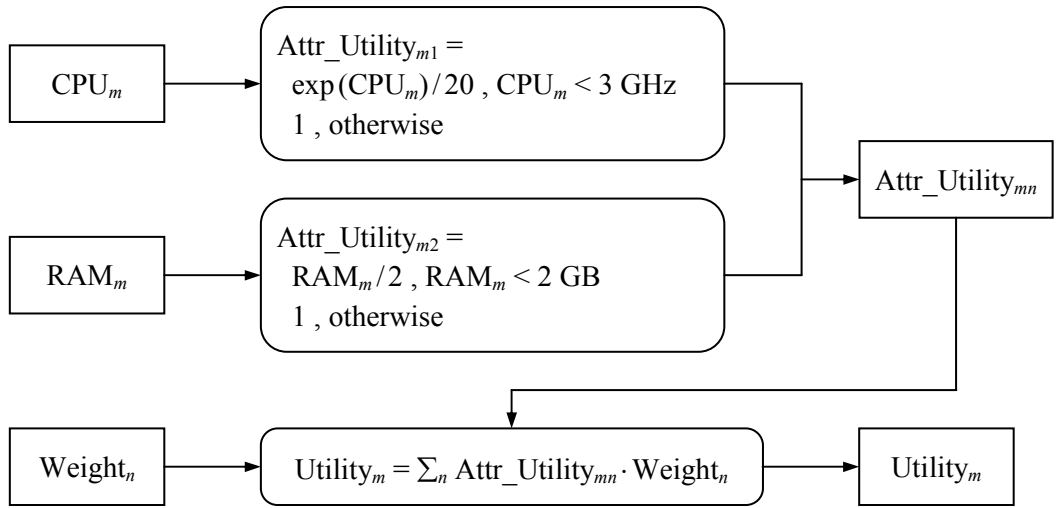

Figure 4: Spreadsheet flow diagram [Ronen et al., 1989] for the weighted additive composition of single-attribute utility functions

Figure 5 gives two logically equivalent, yet physically different, implementations of the multiple-attribute choice problem, which is modelled on Figure 4. The first cascade consists of only five cells and has substantially lower bottom-line error rate than the second cascade, which consists of nine cells. However, all formulas in the latter are very simple and easily understandable, while the former contains a fairly complex formula in the cell labeled 'Utility'. This formula includes six cell references (two cells – 'CPU' and 'RAM' – are referenced twice), several operands and operators, and nests tokens on three levels. Such formulas have the potential to be considerably more liable to errors than simple formulas or cells containing data. Consequently, the cell error rate $e$ should be adjusted in accordance with the complexity of each individual cell. Only then could the reliability of different cascades linking together many cells with simple formulas, or few cells with complex formulas, be reasonably estimated. Hence, it is essential to define and to apply some appropriate measures of complexity.

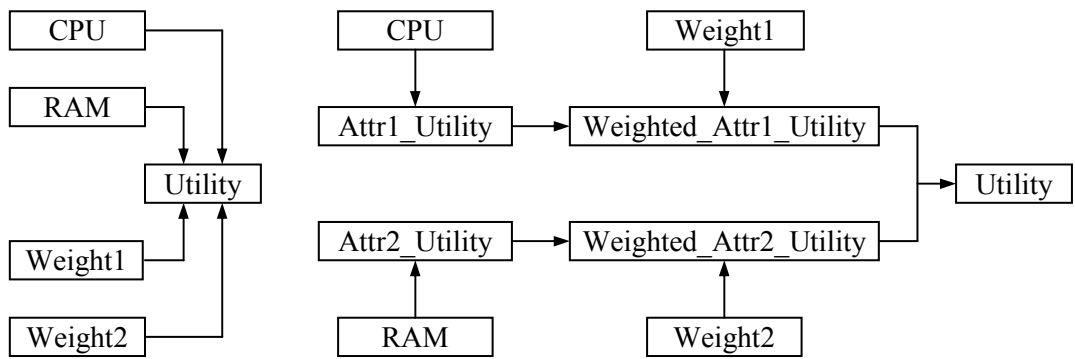

$e = 0.02$, $n = 5 \rightarrow E = 0.0961$, $n = 9 \rightarrow E = 0.1663$

Figure 5: Cell cascades which implement the weighted additive composition of single-attribute utility functions

## 4. CONDITIONAL CONSTRUCTS

It is sometimes useful that a model is allowed to perform different operations dependent upon different values of input variables or intermediate calculations. Yet, logic structure is bound to logical tests within individual cells of the electronic spreadsheet. Conditional functions may thus quickly become overly complex, because conditions have to be nested on many levels within formulas. The complexity additionally increases when sequences of several computational operations are required to be calculated as a consequence of all, or at least some, logical expressions. In the latter case, each operation is usually included on a separate level of a common conditional formula. It may therefore be concluded that auditability, modifiability, reliability and comprehensibility are hindered.

Several approaches to applying complex branches exist. Firstly, conditional statements can be written in the integrated structured and/or object-oriented programming language, as is for example Visual Basic (in the case of Microsoft Excel). More efficient branching of multiple subordinate IF statements is thus enabled, in the sense of better modifiability and comprehensibility. It also allows for the inclusion of an arbitrary number of ELSE IF blocks on the same nesting level. The drawbacks of this approach are that a single cell might potentially contain a lot of references to other cells, and that modellers, auditors and maintainers have to deal with hybrid spreadsheet contents. Hence, it could be argued that neither the auditability nor reliability improve. The second solution is the use of the LOOKUP function [Read and Batson, 1999] which handles arbitrarily many unnested conditions. However, this construct requires comparisons based on logical equivalence. The reason is the origin of the LOOKUP function which is a form of an SQL query. The *condition block* is therefore proposed. It is a slight modification of LOOKUP. The first column (row) of the block declares conditions, and the second column (row) specifies bottom-line operations of computational cell paths. The result of the evaluated condition block is obtained by the *CONDITIONB(range)* function. No lookup value is required, since the value of a cascade belonging to the positively assessed condition is returned.

Finally, the classical IF function may be used in the sense of the *reference branching condition cell*. Proper use of IF would declare two "forward" references corresponding to the positively and negatively assessed condition, respectively. The function thus evaluates the expression and forces one of two cell cascades to execute. It returns a result which is computed by a cascade of operations belonging to the executed branch. In this way, the obtained result can be referenced in other cells regardless of a branch that generated it. Because a "forward reference" cell, or any cell that is directly/indirectly referenced by it, is allowed to contain an analogous logical test, sequencing and nesting of conditions is enabled. The problem with this approach is that only one of two referenced cell cascades returns a relevant, logically reasonable result, while the other, which does not satisfy the imposed expression, evaluates itself to a nonsensical value and might even cause a fault. A possible solution to controlling the execution stream of complex branches is to hide paths. Only paths belonging to the positively evaluated conditions should be normally visible to spreadsheet users. Cells in other paths should be shaded in a predefined way, so that branches are indicated, but not calculated. A similar problem exists for the condition block. It might be solved in the same way.

In order to determine the complexity of logical structure, possible logical paths through a model are considered. Conditional constructs without dependent conditionals are found, because all sub-branches join in cells of these constructs. The complexity of a conditional construct is then recursively computed by summing complexities of all precedent/nested conditional functions, and by adding the number of conditionless computational cascades:

$$O(S) = \left( \sum_{i=1..M} O(S_i) + N \right)^{1+\beta}.$$

Here, $O(S)$ is the complexity of the observed conditional construct $S$, $M$ is the number of nested or precedent conditionals, which are denoted with $S_i$ and are directly or indirectly accessed by the construct $S$, $O(S_i)$ are the complexities of these conditionals, and $N$ is the number of computational cascades not containing any logical tests. When conditional expressions or cascades of operations in an IF function, a reference branching condition cell or a condition block directly/indirectly depend on other conditional constructs, these constructs determine the number of possible logical branches, out of which only some are executed. Otherwise, each computational cascade, not containing any logical tests, adds a single possible branch. If the constant $\beta$ is set to 0, the $O(S)$ measure returns the number of logically disjunctive branches leading to the same bottom-line cell. If $\beta$ has an order of magnitude of $10^{-1}$, two other complexity factors are dealt with.

1. Perceived complexity noticeably increases when conditional functions are nested on many levels of computational cascades, because auditability, modifiability, reliability and comprehensibility tend to decrease. So, the more conditionals there are between any two conditional constructs – the observed and the final one, the more should the observed construct intensify the overall complexity.
2. Different types of conditional constructs – the classical IF function, the reference branching condition cell, and the condition block – should not equally influence the overall complexity. Therefore, a specific value of $\beta$ will have to be experimentally chosen for each individual type of conditionals. In parallel, the degree of risk that is associated with the three different types of constructs will be determined.

Complexity $O(S)$ is measured for each computational cascade separately. It considers all logically disjunctive branches through a cascade. The complexity of an individual branch is computed by applying metrics that were defined in Section 2.

## 5. CONCLUSION

Several spreadsheet complexity metrics were discussed in the paper. These metrics have the purpose of identifying cells which are specially liable to errors, and of adjusting error rates of individual cells in order to properly estimate the probability of wrong bottom-line results. It has to be stressed, though, that the paper presents work in progress, so much effort remains to be invested into further research activities.

The proposed metrics will be validated. It will have to be determined which among them are useful and which must be modified or substituted with more appropriate measurement instruments. A very important task is to apply metrics to actual spreadsheet models in order to determine to what extent the identified complexity factors contribute to obtained error rates. Hence, a correlation between complexity metrics and CER must be specified. Additional complexity factors, which have been overlooked throughout the first research stage, will have to be defined as well. Last, but not least, the proposed metrics should be implemented as a part of an automated analysis tool.

In the paper, only product metrics were dealt with. Traditional software engineering also applies process metrics, such as the development duration and effort. It would make sense to define this kind of quantitative measurements for the purpose of spreadsheet development, and to formally correlate them to the discussed cell complexity factors.